\documentclass[pre,twocolumn,amsmath,amssymb,superscriptaddress,longbibliography,showpacs]{revtex4-1}

\usepackage{graphicx,color}
\usepackage{amsmath,amssymb,latexsym}

\newcommand{\C}{{\mathcal C}}
\newcommand{\mean}[1]{{\left< #1 \right>}}

\begin{document}

\title{Role of trapping and crowding as sources of negative differential mobility}

\author{Marco Baiesi}
\email{baiesi@pd.infn.it}
\affiliation{
Department of Physics and Astronomy, University of Padova, 
Via Marzolo 8, I-35131 Padova, Italy
}
\affiliation{
INFN, Sezione di Padova, Via Marzolo 8, I-35131 Padova, Italy
}

\author{Attilio L. Stella}
\affiliation{
Department of Physics and Astronomy, University of Padova, 
Via Marzolo 8, I-35131 Padova, Italy
}
\affiliation{
INFN, Sezione di Padova, Via Marzolo 8, I-35131 Padova, Italy
}

\author{Carlo Vanderzande}
\affiliation{Faculteit Wetenschappen, Hasselt University, 
3590 Diepenbeek, Belgium}

\affiliation{Instituut theoretische fysica, 
Katholieke Universiteit Leuven, 3001 Leuven, Belgium
}

\date{\today}

\begin{abstract}
Increasing the crowding in an environment does not necessarily trigger
negative differential mobility of strongly pushed particles.
Moreover, the choice of the model, in particular the kind of microscopic jump rates,
may be very relevant in determining the mobility. 
We support these points via simple examples and
we therefore address recent claims saying that crowding in an environment
is likely to promote negative differential mobility.
Trapping of tagged particles enhanced by increasing the force remains
the mechanism determining a drift velocity not monotonous in the driving force.
\end{abstract}

\pacs{ 
05.40.-a,        
05.70.Ln         
}

\maketitle

\section{Introduction.}

In a system at thermodynamic equilibrium, a particle responds to a 
gentle push by moving on average toward the direction of the applied
force. Therefore the mean-velocity to force ratio, called mobility, 
is a positive quantity.
In general, linear response coefficients are positive in equilibrium
because the structure of the fluctuation response
relations involve positive correlations between observables and the 
entropy that perturbations would produce~\cite{tod92}.
Far from equilibrium there is more variability. For example,
thermal conductivities that are not monotonous in the external
forcing have been found~\cite{li06}.

If a pushed particle starts to go slower by increasing the force we have
 the phenomenon of negative differential mobility 
(NDM)~\cite{dha84,zia02,jac08,sel08,lei13,bae13,bas14,ben14,mic15,osh05}
If the particle responds by drifting against 
the force, we speak of absolute negative mobility. 
The latter behavior has been observed
in models subject to ratchet effects~\cite{eic02}, 
in systems driven by periodic forces~\cite{mac07,mar09} and 
for driven Janus particles in corrugated channels~\cite{gho14}.

Here we focus on NDM.
Negative differential conductivity was discovered in the 60's and 
realized in Gunn diodes~\cite{con70} and other semiconductors
at low temperature~\cite{con70,nav76,ala88}. Other examples
of NDM are listed in Ref.~\cite{ben14}. A recent application
of NDM is the sorting of soft matter colloidal particles~\cite{eic10}.

For common overdamped diffusion, NDM is expected when particles' 
surrounding environment contains shallow barriers, called ``traps'' 
hereafter, in which the particle driven by an external force might spend 
more and more time upon an increase of the force. For this reason, the 
mobility of the particle may actually decrease rather than increase if the
force is raised.
An example of this effect was given by particles diffusing within
a percolating cluster~\cite{dha84} 
(an instance of hopping process in disordered media~\cite{van81,bot82}), 
in which traps are represented by the dangling 
ends and all portions with concavity against the direction of the force.
A pedagogical exemplification of this ``getting more from pushing less''
phenomenon was then given by Zia and coworkers~\cite{zia02}. 
In a two-lanes jump system that can be solved exactly, NDM 
emerges when one lane contains a hook that  can trap the particle.
The NDM may be described~\cite{bae13,bas14,mic15} from 
the point of view of a recently developed
linear response theory for general nonequilibrium 
systems~\cite{lip05,bai09,bai11}.
In this context, NDM emerging in simulations
of ring polymers impaled by rods of a lattice with defects~\cite{mic15}
might explain irregular migration speeds in 
experiments comparing linearized and circular DNA plasmids~\cite{mic77,lev87}. 
NDM was also found for particles diffusing in a 
crowded environment~\cite{lei13,bas14,ben14}.

One expects to observe NDM if there is a form
of trapping due to a specific coupling between the tracers (tagged 
particle, polymer, etc.) and their environment.
Jamming of particles, such as in kinetically constrained
models~\cite{jac08,sel08}, may also be seen as a form of mutual
trapping. However, just a density of obstacles does not say 
anything straightforward about such propensity to meet traps if the
features of the obstacles and of the tracers are not explicitly 
specified. Yet,
using the formalism of mesoscopic jump systems, it was recently
proposed~\cite{ben14} that NDM should be a generic feature for particles 
strongly pulled in crowded environments. 

We believe that shifting
the focus from the trapping mechanism to the role of crowding is 
not functional to a better understanding of NDM. 
To support our point, in this paper we show that some of
the results presented in~\cite{ben14}, and more in general the
appearance of NDM, may depend crucially  on the choice 
of microscopic jump rates. Moreover, we use some simple models
to show that crowding may be uncorrelated with NDM or even
anti-correlated. This occurs when too many tagged particles
are introduced in the system, so that traps are saturated.

\section{Jump rates and a basic example.}

As in previous studies, the results are exposed in the context of 
Markov jump processes.
The states of the system are discrete and evolve
with jumps taking place in continuous time. More specifically,
we have in mind subsets of the square lattice where the particle
may occupy free sites and jump to the first neighbors if the transitions
are allowed. Forbidden transition may occur if the target of the jump is
already occupied or if there is an idealized wall separating the two
sites involved in the jump. We use a force $F$ parallel to the 
$x$-axis and periodic boundary conditions in this direction. Hence,
we monitor the average velocity $v \equiv \mean{v}_F$ in the $x$ 
direction.
The differential mobility $\mu \equiv d v / d F$ becomes negative
in the regions where $v$ decreases for increasing $F$.

For the sake of simplicity we use a unit temperature times Boltzmann 
constant combination, $k_B T=1$, and unit spacing between sites of
 the square lattice.
This means that the principle of local detailed balance~\cite{kat83}
is met if the jump rate $k(\C\to \C')$ is related to the jump rate
$k(\C'\to \C)$  of the reversed transition by
\begin{equation}
\frac{k(\C\to \C')}{k(\C'\to \C)} = \exp[U(\C) - U(\C')]
\end{equation}
where $U(\C)$ is the energy associated to the configuration $\C$.
When a particle is subject to a constant force $F$ and the jump 
$\C\to\C'$ is in the direction of the force, the energy difference
becomes the work done by the force, $U(\C) - U(\C') = F$.
When the jump is orthogonal to $F$, simply $U(\C)=U(\C')$. There are
many (actually infinite) choices of the jump rates that satisfy the
local detailed balance condition.

\begin{table}[!b]
\begin{center}
\begin{ruledtabular}
\begin{tabular}{ l  l  l  l}
direction & rate & Model A & Model B \\
\hline
right & $k^+$ &$e^{F/2}/X \sim (1-\epsilon^2)/2$ & $e^{F/2}/Z \sim 1-2 \epsilon$\\
left & $k^-$ & $e^{-F/2}/X \sim \epsilon^2/2$ & $e^{-F/2}/Z \sim \epsilon^2$\\
up/down & $k^0$ & $1/4$ & $1/Z \sim \epsilon$
\end{tabular}
\end{ruledtabular}
\end{center}
\caption{Jump rates of the particle on the square lattice
 with periodic boundary conditions. Here
$X=2(e^{F/2}+e^{-F/2})$  and $Z=e^{F/2}+e^{-F/2}+2$, 
so that all rates $=1/4$ for $F=0$.}
\label{tab:1}
\end{table}

Normalizing jump rates of all four 
exit directions from a site with a common denominator is a popular
{\em choice}  (see e.g.~\cite{zia02,lei13,ben14}), possibly because 
it is elegant and resembles a partition function normalization of Boltzmann weights.
As a consequence, the rate of jumps directed transversely to a large force $F$ 
scale as $\sim \exp(-F)$. 
Another possibility would be to keep transversal rates 
independent of $F$~\cite{bas14}. 
We call {\em model A} the latter and {\em model B} the former. 
The details of these rates are given in Table~\ref{tab:1}, where we also
specify the scaling for large $F$ as a function of $\epsilon = e^{-F/2}$. 
We pick up just these two variants among infinite ones because they
already represent two distinct classes of dynamical behavior.

\begin{figure}[!t]
\includegraphics[width=.9\columnwidth]{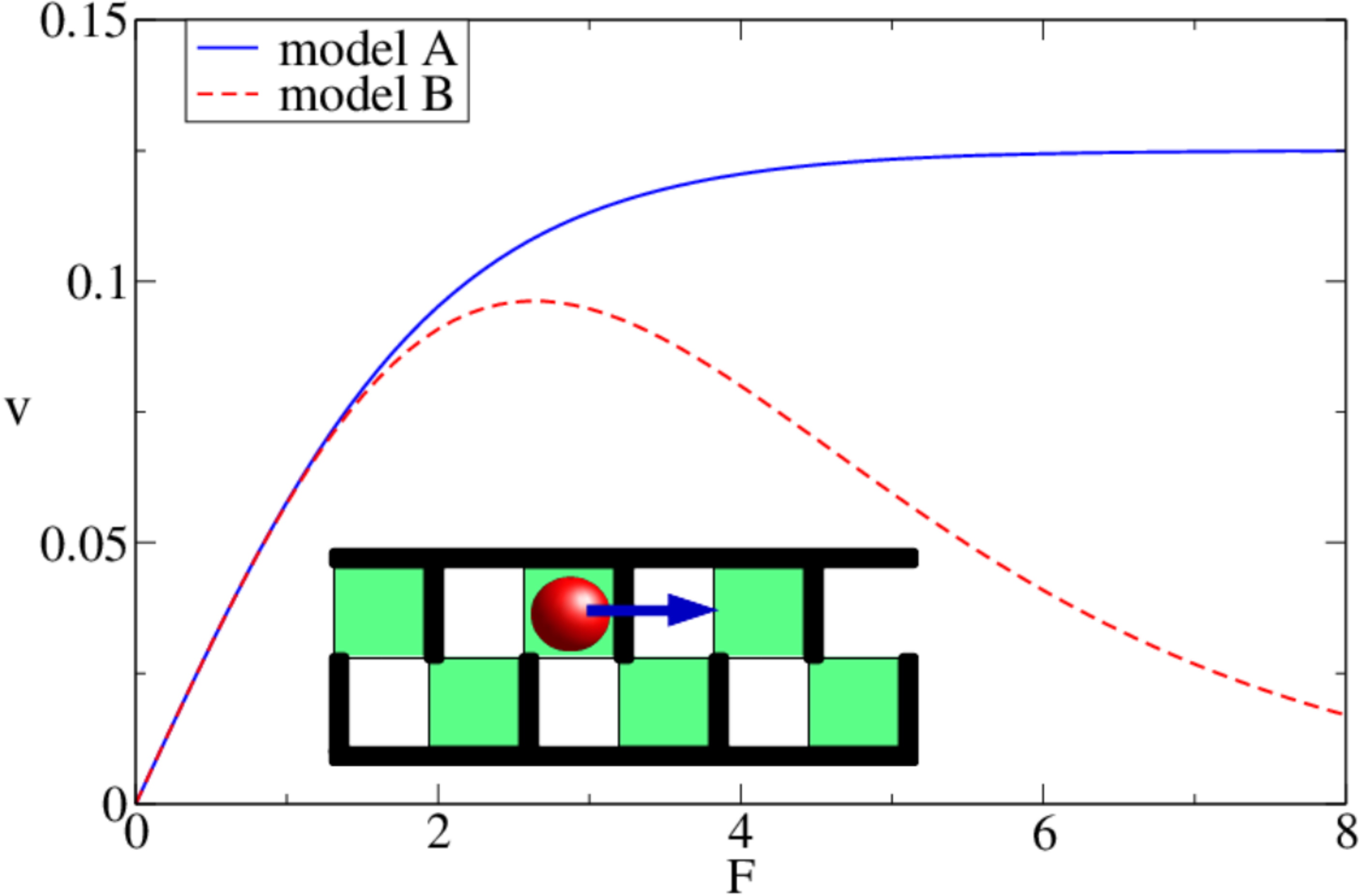}
\caption{(Color online) Sketch of the two-lanes toy model, and its mean velocity 
[see (\ref{velAB})] as
a function of the force, for model A and model B.}
\label{fig:toy}
\end{figure}

In order to have a quick taste of the qualitatively dissimilar mobilities
that model A and B may give,
let us introduce a minimal two-lanes model of particle diffusing
in a meandering channel. This model includes two kind of states: 
$\C_1$ states where the force pushes the particle 
against a barrier (colored boxes in Fig~\ref{fig:toy}) 
and $\C_2$ states where the force pulls the particle away from
a barrier (white boxes). One easily finds the
density of states $p$ and the mean velocity of particles
in the steady state,
\begin{align}
& p(\C_1) =  k^+ + k^0,  \nonumber \\
& p(\C_2) =  k^- + k^0, & v = & k^+ p(\C_2) - k^- p(\C_1). 
\end{align}
 For models A and B the mean velocities are
\begin{equation}
v_{\rm A} = \frac 1 8 \tanh \frac F 2,\qquad
v_{\rm B} = \frac{e^{F/2} - e^{-F/2}}{(e^{F/2} + e^{-F/2} +2)^2},
\label{velAB}
\end{equation}
respectively. The first is clearly increasing with 
$F$ while $v_{\rm B}$ cannot be monotonous 
in $F$ as $v_{\rm B}\sim \epsilon$ for $F\to \infty$
(Fig.~\ref{fig:toy} shows the plots of both functions). 
Hence, Model A does not yield NDM in this case, but Model B does.

A similar fact pertains to scaling theories
focused on the density of obstacles~\cite{lei13,ben14}.
For example, the theory presented in Ref.~\cite{ben14} 
does not yield NDM if constant transversal 
rates are plugged in their Eq.~(2) for the average velocity of the 
tracer, rather than those $\sim \epsilon$ of model B. 
Indeed, with model A rates,  the same formula becomes a 
monotonously increasing function of $F$ 
and  NDM disappears. 

In the following section we show a similar effect for a variant of the
two-dimensional diffusion within slowly moving barriers, which was 
recently used~\cite{bas14,ben14} as a main benchmark for testing ideas on
NDM.

\section{Single tagged particle within floating barriers.}

\begin{figure}[!t]
\begin{tabular}{r r}
(a) & (b)\\
\includegraphics[width=.36\columnwidth]{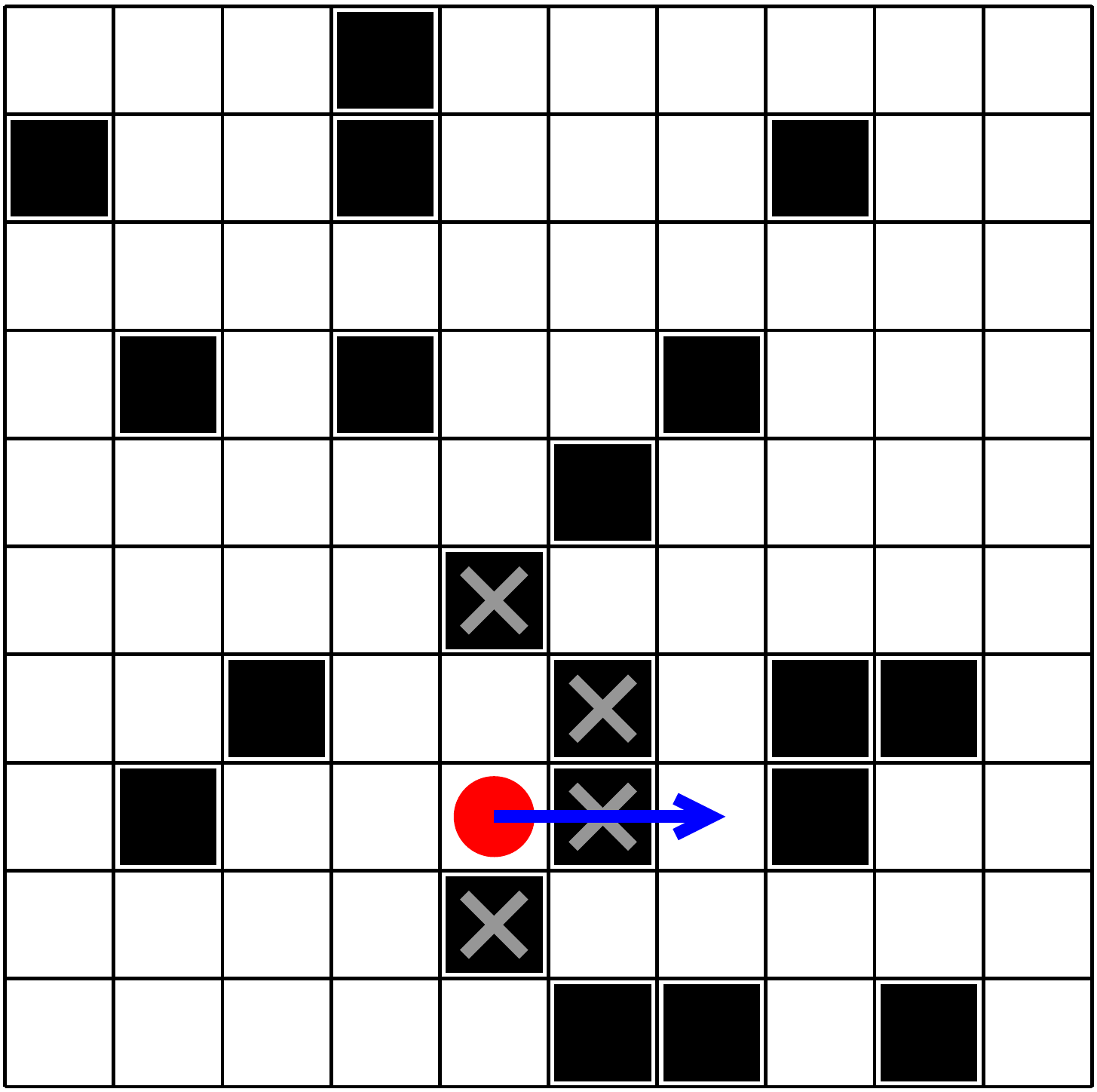}
&
\includegraphics[width=.36\columnwidth]{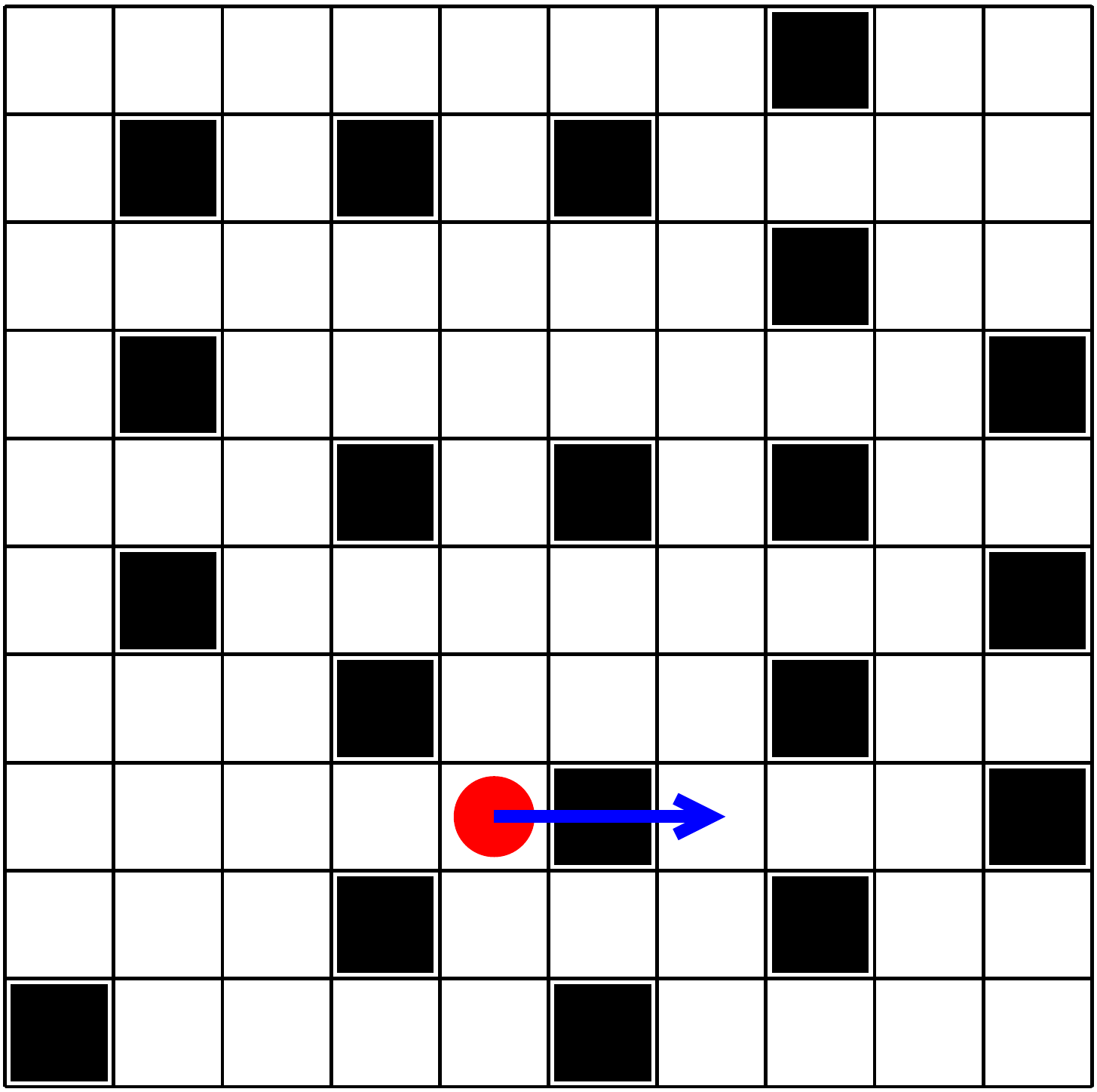}
\end{tabular}
\caption{(Color online) (a) Configuration with density of obstacles $\rho_b=0.2$. The circle 
(red online) is the tagged particle and the squares are the barriers. The four obstacles
marked with a cross are forming a trap for the tracer. (b) Same $\rho_b=0.2$,
when obstacles cannot have vertices overlapped. 
}
\label{fig:ab}
\end{figure}

Model~A was used~\cite{bas14} 
(with a slightly different normalization of rates) 
to show how NDM arises for particles moving in crowded environments 
when the (non-overlapping) obstacles 
have low enough mobility $\gamma$:
given a configuration $\C$ with $N$ obstacles and $1$ tagged particle
[see an example in Fig.~\ref{fig:ab}(a)],
the bare escape rate from $\C$ is 
$\lambda(\C)=\lambda_1 + N \gamma$, where 
$\lambda_1 = k^+ + k^- + 2 k^0$ is the escape rate
of the particle from any site within an empty grid
($\lambda_1=1$ for the rates summarized in Table~\ref{tab:1}).
We extract a waiting time $\tau$ according to an exponential distribution 
$p(\tau)\sim \exp[\lambda(\C) \tau]$. 
After the extracted $\tau$, a move is attempted, picking
up the tagged particle with probability $1/\lambda$ or a random barrier
with probability $\gamma/\lambda$. The tagged particle is then moved with
a random selection of the direction that follows the rates previously 
exposed, while a barrier tries to jump with probability $1/4$ to a random
nearest neighbor. An update realizes the new configuration only
if it is allowed, otherwise the configuration is unchanged
and another waiting time is added to the already spent 
$\tau$ until a valid new configuration is achieved. Such modified
Gillespie algorithm~\cite{gil77} 
is very simple to implement and allows us to sample 
efficiently the dynamics of these models. Note that a time scale
$\tau^*=1/\gamma$ is associated to barriers' motion, as opposed to
a time scale equal to $1$ for the particle motion.

We have performed simulations within square lattices of side $L$ filled
with density $\rho_b = N/L^2$ of floating barriers. An example of
a typical configuration for $\rho_b=0.2$ is shown in
Fig.~\ref{fig:ab}(a). In all simulation 
we find that both model A and B display a similar NDM at sufficiently
large values of $F$. 

\begin{figure}[!t]
\includegraphics[width=.9\columnwidth]{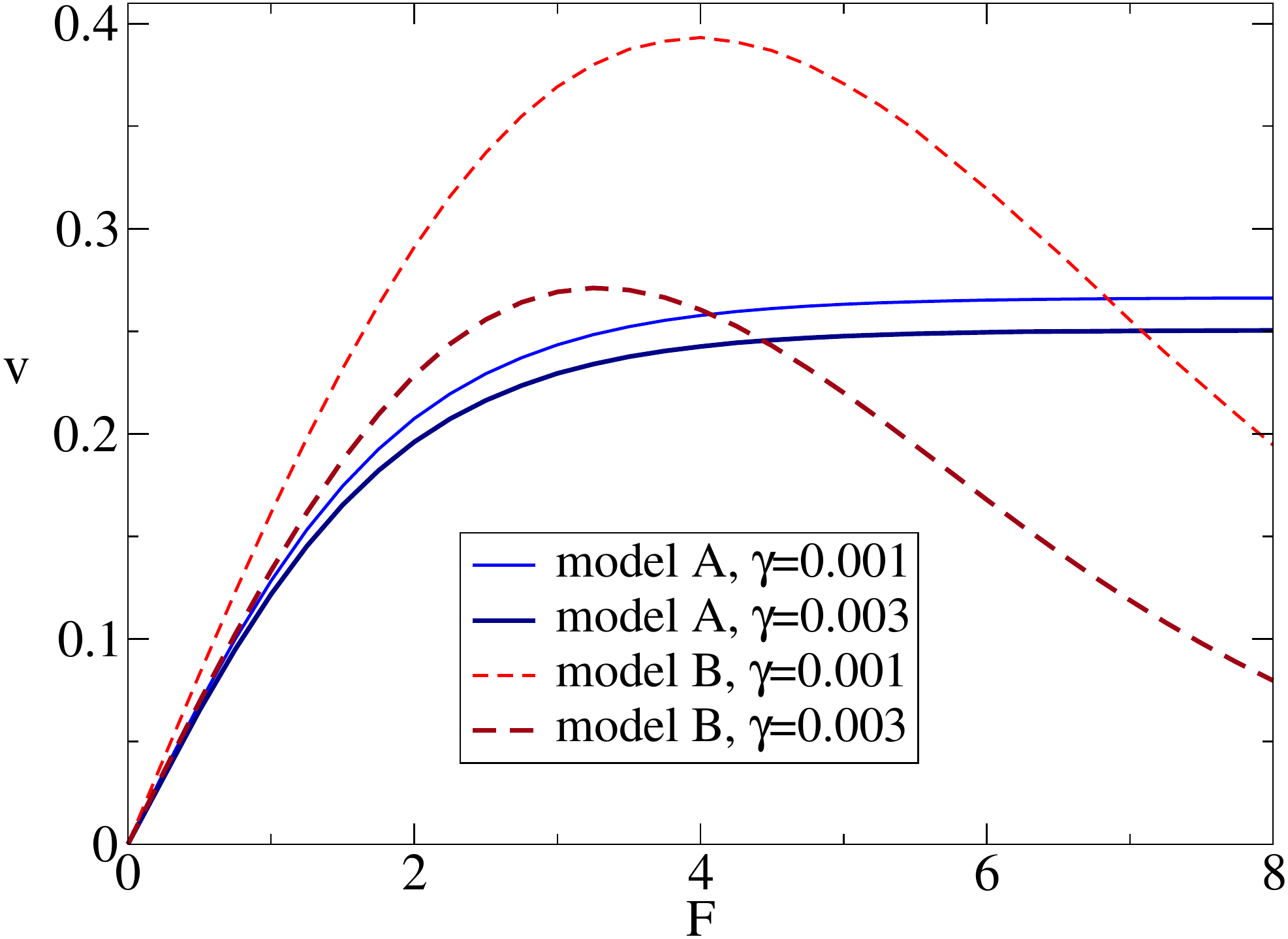}
\caption{(Color online) 
Mean velocity vs.~force for models A and B with non-overlapping corners 
[case (b) of Fig.~\ref{fig:ab}], for $L=20$, $\rho_b=0.2$, and two values
of the obstacles' mobility $\gamma$.
Similar results are found for other small $\gamma$'s, in particular
model A is quite insensitive to the value of $\gamma$.
}
\label{fig:v}
\end{figure}

However, it is not 
possible to switch with ease from one choice of microscopic jump rates to another.
Indeed, even with the same crowding of obstacles, the situation 
changes drastically if these cannot share corners [Fig.~\ref{fig:ab}(b)]
and thus cannot join to form concave traps. 
In this case model A looses NDM while model B continues
to display NDM (Fig.~\ref{fig:v})
because the tracer cannot easily move away from a 
situation as in Fig.~\ref{fig:ab}(b), 
where the rates of model B lead to an effective trapping of the particle.
By now this should be hardly surprising, as
the constraint of non-overlapping barrier's corners
leads to morphologies analogous to those exemplified by
the meandering channel model of the previous section, where
no concavity is present and model A finds no traps to generate NDM.
In the limit of low $\rho_b$ the corner constraint
should be irrelevant. Yet,
theoretical approaches~\cite{lei13,ben14} were used to 
compute a nontrivial mobility also in the limit of low obstacle
densities. Our results suggest that this procedure leads to
an incomplete picture if only model B is considered.

\section{Multiple tagged particles within floating barriers.}

The crowding of an environment may result from the presence of 
different molecular species, and in general it is not straightforward
to anticipate the mobility of particles in complex conditions.
Here we may think of the supercrowded environment of the cell~\cite{mcg10} which affects the diffusive properties of passive particles~\cite{hoe13}. Much less is known about the behaviour of active particles like molecular motors in such an environment~\cite{goy14}. 

As a first step toward more complex systems,
in the simple model described in the previous section
we may think the crowding not only determined by its barriers but also by
a finite, possibly large number of tracers. The total density of occupied sites in the 
system is thus the sum of the density of barriers $\rho_b$ and
the density of tracer particles $\rho_p$. The resulting model is, more than that
of~\cite{ben14}, a combination of the symmetric and asymmetric exclusion processes.
The condition of a finite yet low density of tracers seems similar
to the condition of low filling of bands of electrons, which leads to
NDM in low temperature semiconductors~\cite{van81,ala88}.

\begin{figure}[!t]
\begin{tabular}{r r}
(a) & (b)\\
\includegraphics[width=.44\columnwidth]{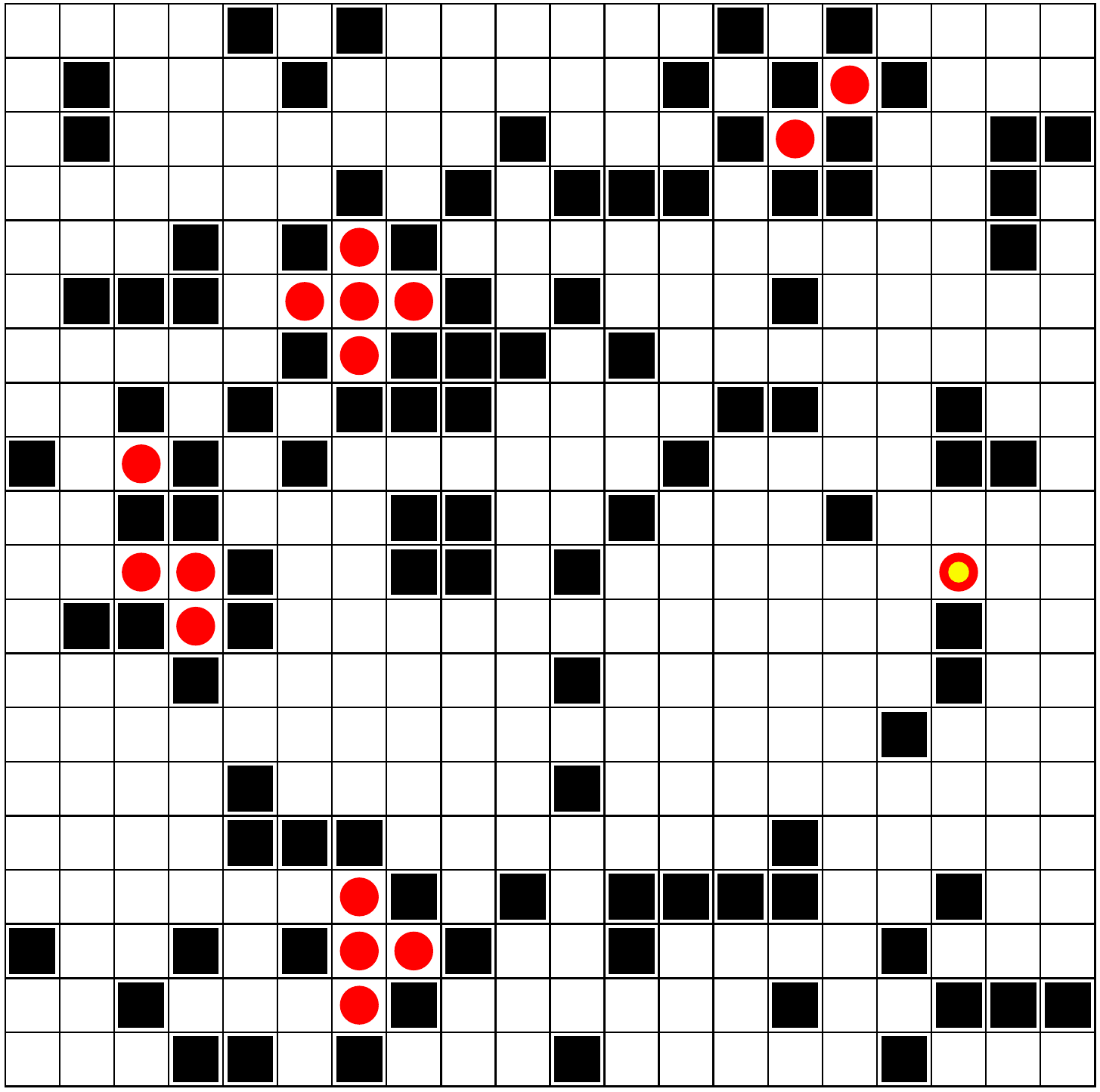}
&
\includegraphics[width=.44\columnwidth]{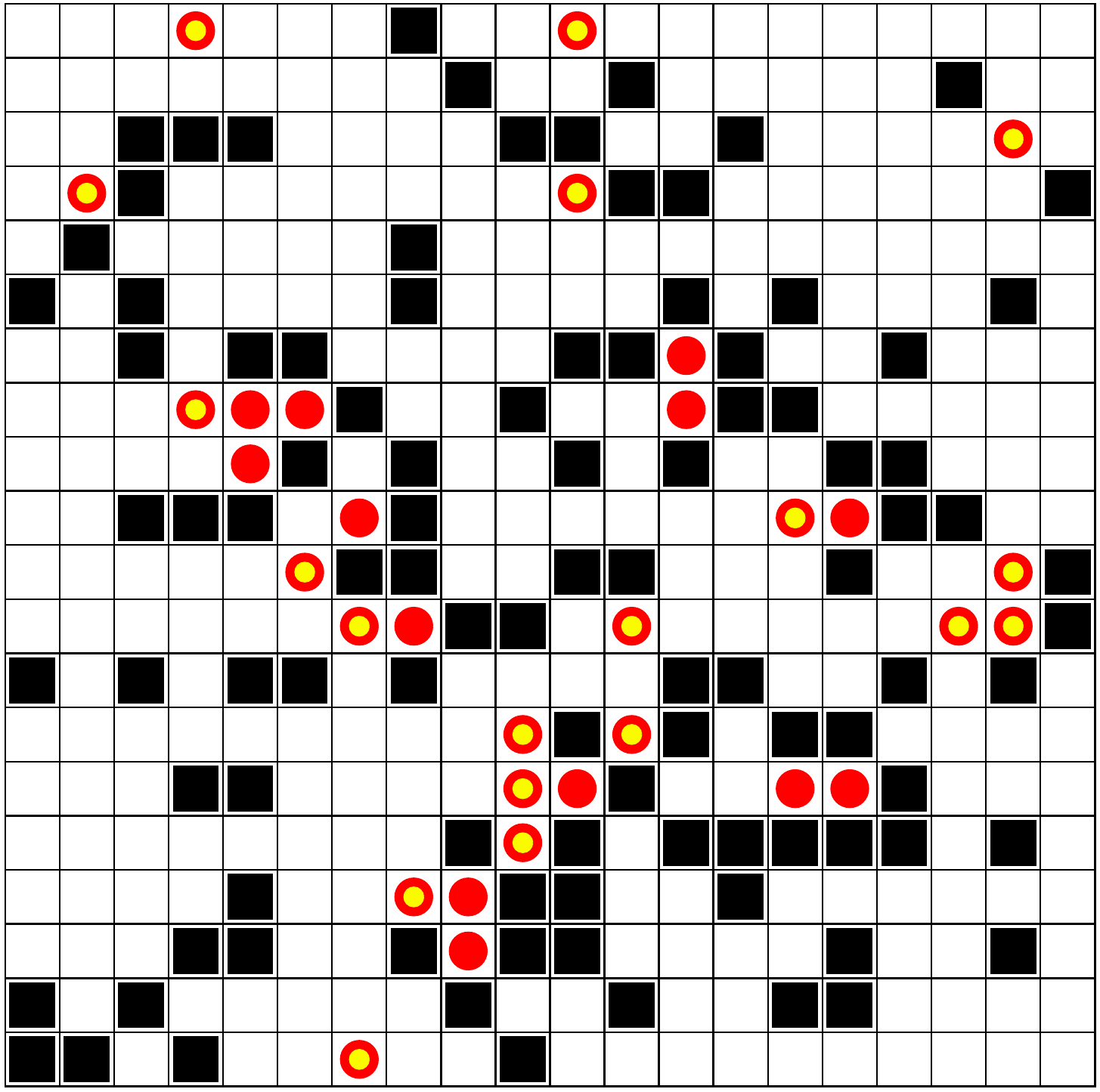}
\end{tabular}
\includegraphics[width=.9\columnwidth]{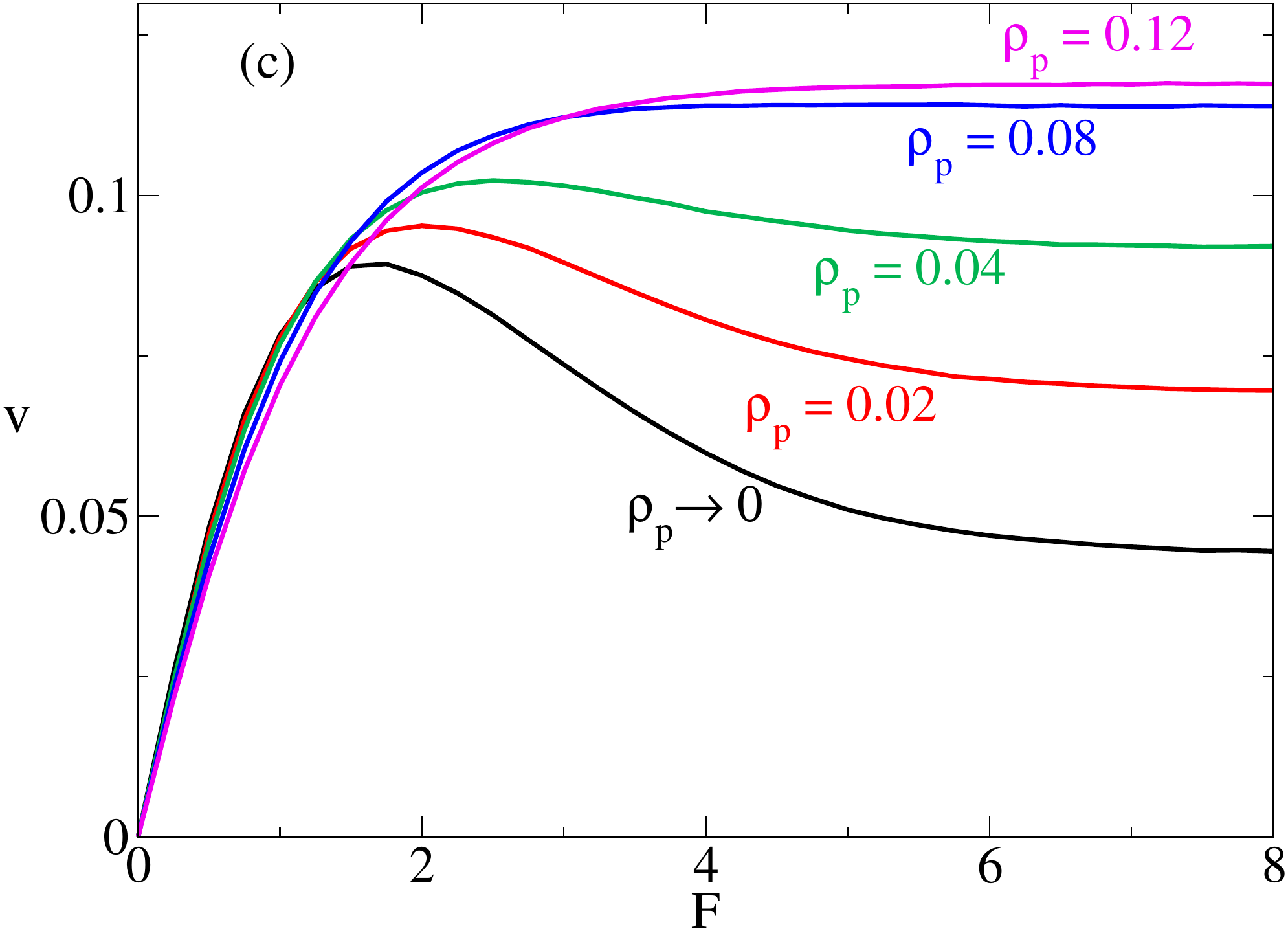}
\caption{(Color online) 
(a) Configuration for model A with $L=20$, $\rho_p=0.04$, 
$\rho_b=0.25$, $\gamma=10^{-3}$, and $F=8$ pushing particles
to the right. 
Tracers that may follow $F$ by eventually never stepping back against it
are drawn as empty circles (only one in this case).
(b) The same for $\rho_p=0.08$.
(c) Mean velocity vs.~force (same parameters) for several particle densities.
}
\label{fig:rho_p}
\end{figure}

For model A it is not difficult to find conditions of low 
$\rho = \rho_p+\rho_b$ where NDM is present as well as conditions
with high $\rho$ where NDM is not observed, see the examples in
Fig.~\ref{fig:rho_p}. The latter state of normal 
mobility arises because the behavior of tracers emerges as an average
of two typical conditions, one of tracers stuck in traps, and one of
tracers free to move because all available traps are occupied by other
tracers. For several values of  $\rho_b$ we find that
the contribution from running tracers determines a normal mobility
for sufficiently high values of $\rho_p$. This is expected to
occur when  there are more particles than the average number of 
traps [e.g.~as in Fig~\ref{fig:rho_p}(b)]. If instead $\rho_p$ is low
enough, trapping is still the dominating behavior 
[see Fig~\ref{fig:rho_p}(a)]

This model therefore is an example where increased crowding 
leads to the disappearance of NDM.
Again, this strengthens our point that the link between crowding 
and the mobility of particles is not straightforward. On the other
hand, the numerical results may be correctly interpreted at least
qualitatively in terms of saturation of trapping effects.
Also in this case one could devise simpler systems where the mechanism
of filling of all the traps is more easily detected. For example,
in the two-lanes models where one lane is occupied by ``hook''
states that act as traps~\cite{zia02,bae13}, 
we have numerically verified that NDM
is present only if the number of tagged particles is at most 
equal to the number of traps.

\section{Discussion and Conclusions.}

We have presented results from some simple models 
suggesting that the phenomenon of NDM in 
many cases can be ascribed to the presence of long living traps that 
catch the strongly pushed traced particles.
There is instead no one to one correspondence between crowding and
NDM. In particular, according to our results, it seems far fetched to 
draw conclusions from either simulations or from mesoscopic theories if these 
rely on the specific choice of the jump rates.

The choice of transversal jump rates normalized so that they 
decrease with the force is quite popular 
but it needs to be understood and justified. 
For example, starting from the concept of Brownian motion of 
a particle in a fluid, we do not see how the increasing force could 
lead to a weaker transversal motion, namely to a diffusion coefficient 
in the fluid that depends on a transversal $F$. Also for particles 
diffusing by jumping in a pattern of energy minima it is not general 
to find a decrease of transversal motion generated by an applied force. 
Think for example of a particle density within a minimum, 
at the microscopic level of description
(while the mesoscopic level is the one of jump systems describing
the energy minimum as a single state).
If the applied force shifts the maximum of this density closer to the 
energetic saddle points that connect to 
nearby states, also in transverse directions, 
we would witness a case where the force would actually enhance 
the transversal motion. It seems that such scenario is in principle
as likely as that where the transversal motion is depressed by the
force. Hence, transversal jump rates that decrease exponentially
with the applied force should be tested in parallel with other choices,
keeping in mind that any specific system might be a realization of one of
these choices.

Our examples also show that focusing on the density of 
objects responsible for the crowding is not
a stand-alone strategy. One needs
first to determine whether the tagged mobile objects (e.g. particles or polymers)
might become more stuck by the crowded environment when the pushing
force is increased. Furthermore, it is also relevant to check 
if on average the traps are at least as many as the tagged particles. The study
of NDM in more complex simulations than those considered so far 
(such as the condition outlined in the previous section where multiple 
tracers coexists) should help further clarifying these issues.
Of course this is just part of a broader scenario, where kinetic
constraints, jamming, glassy dynamics, alternating external forces,
self-propulsion, etc.~furnish many more mechanisms leading to
negative mobility.


%

\end{document}